\pdfoutput=1
\documentclass[aps,twocolumn,prb,groupedaddress]{revtex4}
\usepackage{graphicx}
\usepackage{amsmath}
\usepackage[T1]{fontenc}
\usepackage{paralist}
\usepackage{verbatim}
\usepackage{float}
\usepackage{placeins}
\usepackage{epstopdf}

\newcommand{\Eq}[1]{Eq.(\ref{#1})}% \Eq{eq:abc}
\newcommand{\Fig}[1]{Fig.\,\ref{#1}}% \Fig{fig:abc}
\newcommand{\Onlinecite}[1]{Ref.\,\onlinecite{#1}} % \Onlinecite{abc}

\newcommand{\be}{\begin{equation}}
\newcommand{\ee}{\end{equation}}
\newcommand{\bea}{\begin{eqnarray}}
\newcommand{\eea}{\end{eqnarray}}

\newcommand{\bk}{\mathbf{k}}
\newcommand{\bkP}{\mathbf{k}_{\rm P}}
\newcommand{\bkS}{\mathbf{k}_{\rm S}}
\newcommand{\bkI}{\mathbf{k}_{\rm I}}

\newcommand{\EP}{E_{\rm P}}
\newcommand{\ES}{E_{\rm S}}
\newcommand{\EI}{E_{\rm I}}
\newcommand{\ELP}{E_{\rm LP}}
\newcommand{\EMP}{E_{\rm MP}}
\newcommand{\EUP}{E_{\rm UP}}

\newcommand{\Ec}{E_{\rm c}}
\newcommand{\Ehh}{E_{\rm hh}}
\newcommand{\Elh}{E_{\rm lh}}

\newcommand{\x}[1]{x_{\rm #1}}
\newcommand{\ghh}{\gamma_{\rm hh}}
\newcommand{\glh}{\gamma_{\rm lh}}
\newcommand{\gc}{\gamma_{\rm c}}

\newcommand{\g}[1]{\gamma_{\rm #1}}
\newcommand{\Ohh}{\Omega_{\rm hh}}
\newcommand{\Olh}{\Omega_{\rm lh}}
\newcommand{\Eh}[1]{\widehat{E}_{\rm #1}}
\newcommand{\Et}[1]{\tilde{E}_{\rm #1}}
\newcommand{\Epm}[2]{\widehat{E}^{#2}_{\rm #1}}
\newcommand{\Eint}[1]{E_{\rm #1}^{\rm int}}
\newcommand{\Eren}[1]{E_{\rm #1}^{\rm ren}}
\newcommand{\PP}{P_{\rm P}}
\newcommand{\MPar}[1]{M^{\rm par}_{\rm #1}}
\newcommand{\IP}[1]{I^{\rm par}_{\rm #1}}

\newcommand{\Dc}{\Delta_{\rm c}}
\newcommand{\As}[1]{A^*_{\rm #1}}
\newcommand{\De}[1]{\Delta_{\rm #1}}
\newcommand{\de}[1]{\delta_{\rm #1}}

\newif\iffig\figtrue

\begin{document}
\textheight 25cm %\preprint{}
\sloppy

\title{Parametric scattering of microcavity polaritons into ghost branches}
\author{Joanna M Zajac}
\email[Electronic address:]{joanna.m.zajac@gmail.com}
\author{Wolfgang Langbein}
\affiliation{School of Physics and Astronomy, Cardiff University,
The Parade, Cardiff CF24 3AA, United Kingdom}

\date{\today}
\begin{abstract}
Polaritons of defined momentum and energy are excited resonantly on the lower polariton branch
of a planar semiconductor microcavity in the strong coupling regime, and the spectrally and
momentum resolved emission is analyzed. We observe ghost branches from scattering within the
lower polariton branch, as well as from scattering to the middle polariton branch, showing the
non-linear mixing between different branches. Extending the theoretical treatment of
spontaneous parametric luminescence developed in Ciuti {\em et al.}, Phys. Rev. B {\bf 63},
041303 (2001), the eigenmodes of the driven polariton system and its photoluminescence are
modeled.
\end{abstract}

\maketitle
\begin{comment}
\end{comment}

Cavity excitons-polaritons in planar semiconductor microcavities are quasi-particles resulting
from strong coupling between the Fabry-P\'{e}rot cavity mode and excitonic resonance of the
semiconductor inside the cavity. Below the exciton saturation density, polaritons can be
treated as composite bosons \cite{Kavokin07microcavities}. They inherit features of exciton
and photon constituents resulting in strong interactions and a in-plane dispersion and
propagation dominated by the cavity mode. The parametric scattering of microcavity polaritons
is described in lowest order by the third-order susceptibility\cite{Boyd08book}. Given a
coherent population of "pump" (P) polaritons, which are scattered into "signal" (S) and
"idler" (I) polaritons, the phase matching in time and space results in the conservation of
energy $2\EP=\ES+\EI$, and momenta $2\bkP=\bkS+\bkI$ where $\bk$ is the wavevector. The
scattering is resonant to the eigenstates of the system, which in the investigated sample are
the polaritons of the lower, middle and upper branches with the energies
$\ELP(\bk),\EMP(\bk),\EUP(\bk)$. This scattering enables optical parametric amplification
\cite{SavvidisPRL00}. A theoretical model describing the spontaneous parametric fluorescence
was discussed in \Onlinecite{CiutiPRB01}, and extended to stimulated emission in
\Onlinecite{CiutiSST03}. Spontaneous parametric emission was experimentally investigated in
\Onlinecite{LangbeinPRB04b,LangbeinPSS05} showing the scattering into the phase-matched
8-shapes in momentum space, which was then shown to provide entangled photon
sources\cite{SavastaPRL05,LangbeinPSS05}, which was recently extended to one-dimensional
cavity structures \cite{AbbarchiPRB11,ArdizzonePSSB12}. In this letter, we report on
spontaneous parametric scattering of resonantly excited polaritons onto real and ghost
branches.

The microcavity sample\cite{BorriPRB01} investigated here
%(see Fig\,\ref{fig:Denmark_sample_design}),
is a 1$\lambda$ Al$_{0.05}$Ga$_{0.95}$As cavity with a single 15\,nm GaAs quantum well with
5\,nm Al$_{0.3}$Ga$_{0.7}$As barriers in its center, providing two excitonic resonances, the
heavy hole and the light hole exciton. The cavity is surrounded by
AlAs/Al$_{0.15}$Ga$_{0.85}$As distributed Bragg reflectors with 25(16) periods on the
bottom(top) of the epilayer. The cavity mode energy gradient was about 1.5\,meV/mm, which
allowed to adjust the detuning between cavity and heavy-hole exciton $\Dc = \Ec-\Ehh$. The use
of a wide binary GaAs well eliminates the alloy disorder found in InGaAs/GaAs quantum
wells\cite{JensenJAP99}, resulting in an inhomogeneous exciton linewidth of\cite{BorriPRB01}
$170\,\mu$eV.
%\begin{figure}
%\iffig\centerline{\includegraphics[width=\columnwidth]{fig/Denmark_sample_design_v2} }\fi
%\caption {Sample design see detailed description in text.} \label{fig:Denmark_sample_design}
%\end{figure}
The Al$_{0.05}$Ga$_{0.95}$As cavity reduces the carrier confinement and thus the carrier
trapping and the related homogenous broadening\cite{JensenAPL00}. The resulting exciton
linewidth was measured here at full-width half maximum as $\ghh=150\,\mu$eV using reflection
spectroscopy. The cavity linewidth $\gc$ of about $300\,\mu$eV is limited by the reflectivity
of the top Bragg mirror.

The sample was mounted in a helium bath cryostat at a temperature of 5\,K and a vapor pressure
of 200\,mbar. To measure the polariton dispersion, we used a weak pulsed excitation with a
mode-locked Ti:Sapphire laser (Coherent Mira) delivering 100\,fs pulses at 76\,MHz repetition
rate and a spectral width of approximately 20\,meV. The excitation was focused to a
diffraction limited spot of $1.5\,\mu$m with a 0.5NA lens having a wavevector range of
$|k|\leq 4/\mu$m. To excite pump polaritons for parametric scattering, we used a linearly
polarized single-mode CW external cavity diode laser with a spectral width of 20\,neV.
Two-dimensional excitation wavevector control was realized by a mirror on a gimbal mount,
which was imaged onto the sample\cite{LangbeinPRB04b}. The beam divergence at the mirror was
adjusted to create a gaussian tail at the sample, providing the minimum wavevector spread for
a given excitation size. The beam diameter at $1/e$ intensity on the sample was $70\,\mu$m,
corresponding to a wavevector spread of $|\bk| \leq 0.09/\mu$m. In order to avoid sample
heating, the excitation was chopped by an acousto-optic modulator producing pulses of
$1\,\mu$s pulse duration at 1\% duty cycle. The peak intensity on the sample $I$ was about
$10^3$\,W/cm$^2$. The resonantly created polariton density
$N_{\rm LP}(\bkP) \simeq I\tau T/\EP \simeq 10^{7}$/cm$^2$\,\footnote{$N_{\rm LP}$ was calculated
using the lower polariton lifetime $\tau =\hbar/\g{LP}=1.5$\,ps with $\g{LP}=400\,\mu$eV,
and the DBR top mirror transmission for $\g{cav} = 300\,\mu$eV was calculated as $T = 0.3$\%}
%for calculation see angles.mws in the same folder
The reciprocal space ($\bk$) of the cross-linearly polarized emission was imaged onto the
input slit of a high resolution ($20\,\mu$eV) imaging spectrometer and detected using a
CCD-Camera \cite{ZajacPRB12,LangbeinRNC10}.

\begin{figure}
\iffig\includegraphics[width=0.9\columnwidth]{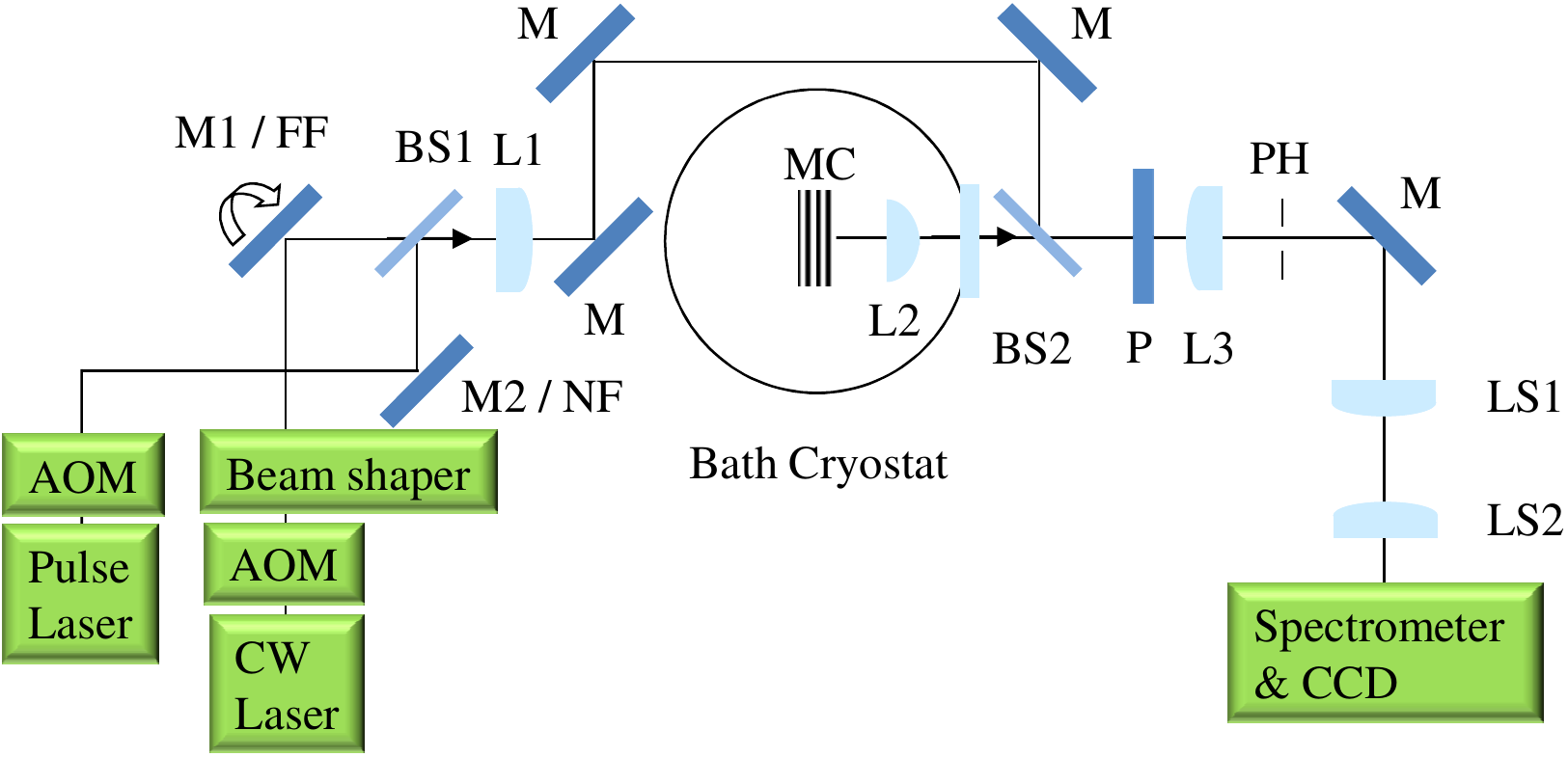}\fi \caption {Optical setup used.
M1,M2: Gimbal mounted mirrors, L1-L6 Lenses, MC: Microcavity sample, LS1,LS2 movable lenses
for imaging, dashed lines: removable mirrors, BS1,BS2 Beam-splitters.} \label{fig:setup_para}
\end{figure}
%
%\section{Results}
%\label{sec:Results}
%
\begin{figure}
\iffig\centerline{ \includegraphics[width=1\columnwidth]{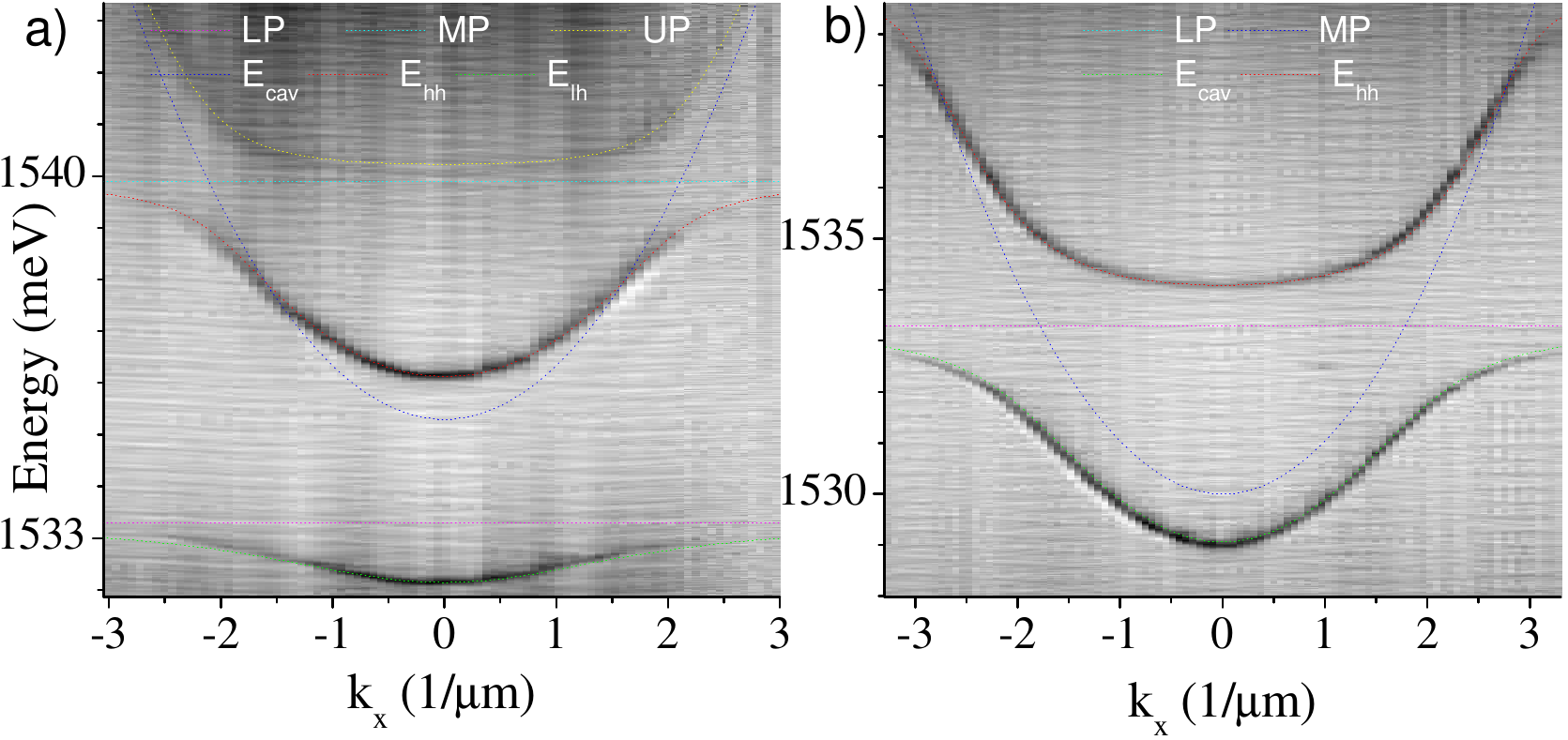} }\fi
\caption{Reflection of the microcavity as function of photon energy and wavevector
$\bk=(k_x,0)$. a) positive detuning $\Dc=5$\,meV. b) negative detuning $\Dc=-4$\,meV. The
calculated polariton dispersions $\Et{B}(\bk)$ are given by lines as labeled.}
\label{fig:UnperturbedDisp}
\end{figure}
The polariton dispersion in the low-intensity regime was measured using $\bk$ resolved
reflection spectroscopy as shown in \Fig{fig:UnperturbedDisp}, and modeled with the coupled
three oscillator model for the cavity mode, heavy- and light-hole exciton\cite{JensenAPL00,
LangbeinJPCM04}. From these fits of $E_{\rm B}(\bk)$ to the set of polariton branches
B=\{LP,MP,UP\}, we deduced the exciton energies $\Ehh=1.5333$\,eV, $\Elh=1.5399$\,eV, and the
Rabi splittings $2\,\Ohh=3.7$\,meV, $2\,\Olh=2.4$\,meV, for heavy- and
light-hole excitons, respectively. The exciton and polariton linewidths of this sample were
previously compared\cite{BorriPRB01} with the linewidth averaging model, in which the
polariton linewidth $\g{B}$ is a weighted average of $\gc$ and the exciton linewidths
$\ghh,\glh$,
\be \g{B} = \x{lh,B}\ghh + \x{lh,B}\glh + c_{\rm B}\gc \label{eq:linewidthAvModel} \ee
with the contents of cavity $c_{\rm B}$, heavy hole exciton $\x{hh,B}$, and light hole exciton
$\x{lh,B}$ in the polariton\cite{LangbeinPRB04a}.  The model is assuming Lorenzian lineshapes,
and shows sufficient agreement\cite{BorriPRB01} with the experiment for the LP at zero and
negative detuning. Increasing the exciton density as relevant in our experiments, the exciton
linewidth is dominated by exciton-exciton scattering, which has a different non-Lorentzian
shape compared to inhomogeneous broadening.
\begin{figure*}
\iffig\includegraphics[width=0.9\textwidth]{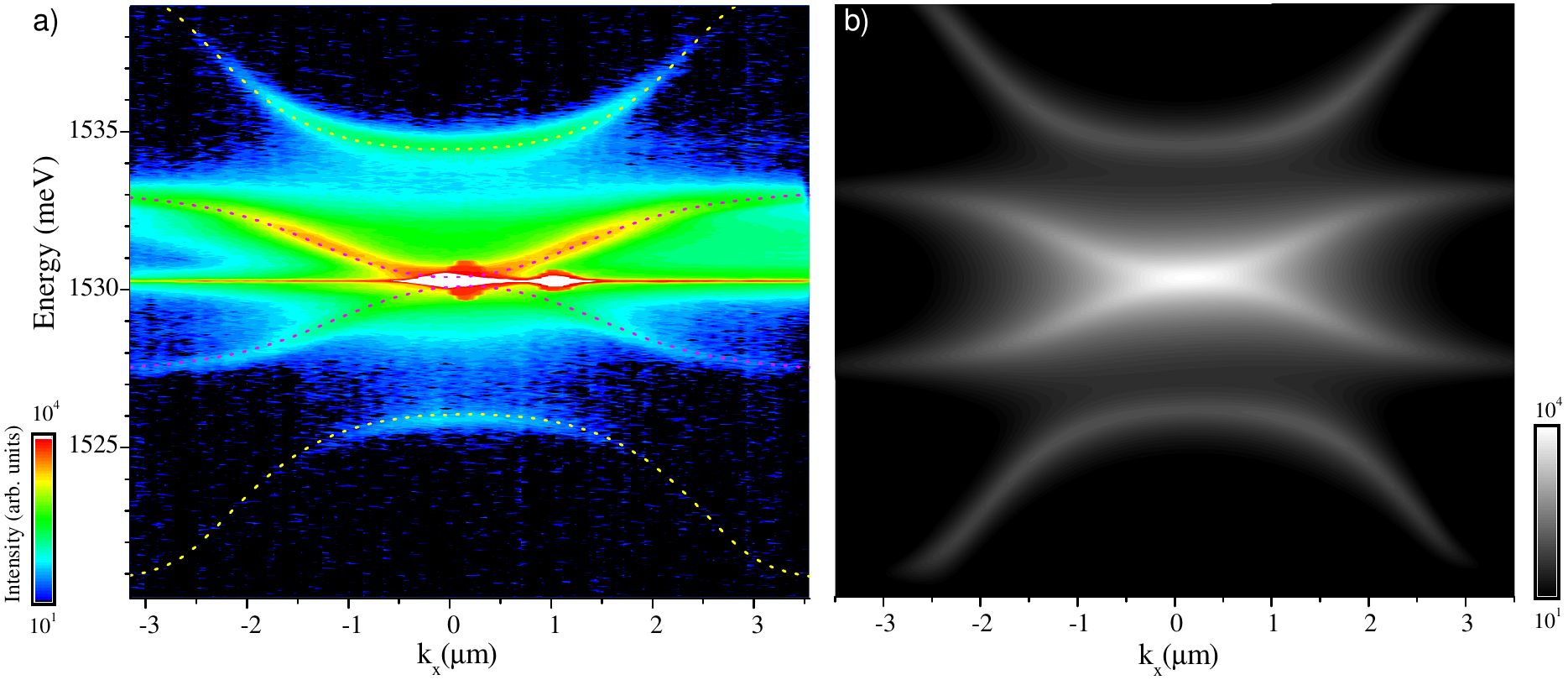}\fi \caption{Energy and
wavevector resolved emission intensity $I(\bk,\omega)$ on a logarithmic scale as indicated,
for a pump energy $\EP=1530.25\,$meV and wavevector $\bkP=(0.1,0)/\mu$m at a cavity detuning
of $\Dc= -1.7$\,meV. a) Measured $I(\bk,\omega)$ for $\bk=(k_x,0.4/\mu\mbox{m})$. Lines:
Eigenmodes $\Epm{B}{\pm}(\bk)$ of \Eq{eq:modMatrixPara}, for B=LP in magenta, and for B=MP in
yellow. b) Parametric emission $\IP{}(\bk,\omega)$ calculated using \Eq{eq:PLLPMP}.}
\label{fig:D120218012P}
\end{figure*}

The parametric emission was modeled following \Onlinecite{CiutiPRB01}, where the polaritons
are excited resonantly with a pump field $\PP(t) = \langle p^{\dag}_{\rm LP}(\bkP,t) \rangle$
of defined wavevector $\bkP$ and photon energy $\EP$ within the LP branch. The polaritons of
signal and idler are coupled by a momentum conserving exciton-exciton scattering proportional
to the pump intensity, described by off-diagonal terms in an anti-hermitian coupling matrix.
In the following we use the renormalised complex polariton energies $\Eh{B} = E_{\rm
B}-i\g{B}+ \Eren{B}|\PP|^2 $.  The polariton-polariton interaction term $\Eren{B}$ was
determined using Eq.\,9 in \Onlinecite{LangbeinPRB04b}
\bea && \Eren{B}(\bk) = 2\x{LP}(\bkP)\x{B}(\bk) \left\{12 E_{\rm X} + \right. \\
&& \nonumber \left. \frac{16\pi}{7} \Ohh \left [ \sqrt{\x{LP}^{-1}(\bkP)-1} +
\sqrt{\x{B}^{-1}(\bk)-1} \right] \right\}, \eea
with an exciton binding energy of $E_{\rm X} = 8$\,meV. The excitonic content $\x{B}$ was
taken as the sum $\x{B} = \x{hh,B} + \x{lh,B}$ of heavy and light hole content. The expression
holds for circular polarization, so that for the cross-linear polarization configuration used
here in the regime where the renormalisation is smaller than the linewidth we expect some
deviations in the overall scattering strength. For higher polariton densities the
spin-dependent interaction is influencing the dynamics significantly\cite{SolnyshkovPRB08}

Neglecting higher-order scattering processes and Langevin terms of the external light field,
the steady-state emission from these branches was derived in an analytical form (Eq.\,9 of
\Onlinecite{CiutiPRB01}) as a function of the steady state population of the signal $N_{\rm
B}({\bkS})=\langle p^{\dag}_{\rm B}(\bkS,0)p_{\rm B}(\bkS,0) \rangle$ and an anomalous
parametric correlation amplitude between signal and idler polaritons $\As{B}(\bkS)=\langle
p^{\dag}_{\rm B}(\bkS,0)p^{\dag}_{\rm B}(\bkI,0) \rangle$ where $p_{\rm B}(\bk,t)$ is the
time-dependent polariton operator of branch B and wavevector $\bk$. We extended the model to
include the middle polariton branch (MP) resulting in a corresponding ghost branch (MP*). The
coupling matrix for the different branches is given by

\be \MPar{B}=\left( \begin{array}{c c}
\Eh{B}(\bkS) & \Eint{B} \PP^2 \\
-(\Eint{B} \PP^2)^* & 2\EP-\Eh{B}^*(\bkI)\\
\end{array} \right) \label{eq:modMatrixPara} \ee
having the eigenvalues $\Epm{B}{\pm}(\bkS)$. The interaction energy $\Eint{B}$ is given by
Eq.\,8 in \Onlinecite{LangbeinPRB04b},

\bea \Eint{B}(\bkS) &=& \x{LP}(\bkP)\sqrt{\x{B}(\bkS)\x{B}(\bkI)} \times \\
&& \nonumber \left\{12 E_{\rm X} +
\frac{16\pi}{7} \Ohh \left [ 2\sqrt{\x{LP}^{-1}(\bkP)-1} +\right.\right. \\
&& \nonumber  \left.\left.\sqrt{\x{B}^{-1}(\bkS)-1} +\sqrt{\x{B}^{-1}(\bkI)-1} \right]
\right\}, \eea

The parametric emission intensity of each polariton branch $\IP{B}$ is then given by
\bea && \IP{B}(\bkS,\omega) \propto c_{\rm B}(\bkS)\times\label{eq:PLLPMP} \\
&& \Im \left \{ \frac{\De{B}(\bkS,\omega) N_{\rm
B}(\bkS)+\Eint{B}(\bkS)\PP^{2}\As{B}(\bkS)}{\left(\Epm{B}{+}(\bkS)-\hbar
\omega\right)\left(\hbar \omega - \Epm{B}{-}(\bkS)\right)} \right\} \nonumber\eea
with
\be \As{B}(\bkS) =
\frac{\Eint{B}(\bkS)\PP^{2}\de{B}(\bkS)}{\left|\de{B}(\bkS)\right|^2-\frac{\left(\g{B}(\bkS)+\g{B}
(\bkI)\right)^2}{\g{B}(\bkS)\g{B}(\bkI)} \left|\Eint{B}(\bkS)\PP^{2}\right|^2}, \nonumber\ee
$ N_{\rm B}(\bkS) = \Im\left\{ \Eint{B}(\bkS) \PP^{2}\As{B}(\bkS) \right\}/\g{B}(\bkS)$
, the emission detuning
$\De{B}(\bkS,\omega) = \hbar\omega + \Eh{B}^*(\bkI) - 2\EP $
and the signal-idler detuning
$\de{B}(\bkS) = 2\EP -\Eh{B}^*(\bkS)-\Eh{B}^*(\bkI)$.
The total emission $\IP{}$ is the sum of $\IP{B}$ over all branches B. This theoretical
treatment is valid below the threshold for parametric oscillation given by the condition $\Im
( \widehat{E}^{-}_{B}(\bkS) )<0$. We used the complex polariton energies $\Eh{B}$ calculated
in the three coupled oscillator model with a k-dependent broadening from
\Eq{eq:linewidthAvModel} with $\gc = 300\mu$eV and $\glh = \ghh = 400\mu$eV. The exciton
linewidths are higher than measured in the low intensity regime, which we attribute to
exciton-exciton scattering by the higher exciton density in the parametric scattering
experiments \cite{HuynhPRL03}. The pump is assumed to be resonant to the LP branch. The
simulations shown are well below the threshold, for which the renormalization is negligible
and $\IP{}$ is independent of the pump intensity up to a scaling factor. We used $P_{\bkP} = 10^{-3}$, $N_{LP}(\bkP)\simeq10^8$/cm$^2$ for exciton Bohr radius $\lambda_X = 14$\,nm\cite{CiutiPRB01}.
Simulations were made with a step size of $20\,\mu$eV in $\hbar\omega$ and $0.06/\mu$m in $\bkS$.
%\subsection{Experimental results and discussion}
%\label{sec:experiment}

We now discuss the measured microcavity emission for different pump energies and wavevectors
together with corresponding results of simulations. We commence with a pump close to the
dispersion minimum at $\EP=1530.15$\,meV and $\bkP=(0.1,0)/\mu$m for a cavity detuning
$\Dc=-1.7$\,meV, shown in \Fig{fig:D120218012P}.
% Two data are binned together proving higher S/N ratio and k-resolution of 0.14.
The measured emission $I((k_x,0.4/\mu\mbox{m}),\omega)$ shown in \Fig{fig:D120218012P}a shows
the dominant emission from the LP and from the pump which is scattered elastically by disorder
towards the detection wavevector range.
%{\bf WL a inset with a 2d $\bk$ plot at fixed energy and the position of pump and crossection
%would help,JZ: origin does not put plots on top of matrices}.
The emission from the MP is about 2 orders of magnitude weaker, and the ghost branches LP* and
MP* are 2-4 orders of magnitude weaker and show a reversed dispersion. The corresponding
predicted eigenvalues $\Epm{B}{\pm}(\bk)$ of \Eq{eq:modMatrixPara} are following the observed
emission peaks.  For a more detailed comparison with theory, we give in \Fig{fig:D120218012P}b
the calculated parametric emission intensity $\IP{}$, which shows a semi-quantitative
agreement with the experimental result. The main deviation is the observed intensity of the
ghost branches, which in the experiment is much weaker than in the simulation. This is
actually expected as the model accounts for radiative broadening only, such that all
parametrically scattered polaritons are emitted, resulting in equal intensities of signal and
idler. In the experiment, a significant part of the broadening at higher $\bk$ is due to the
exciton linewidth (see \Eq{eq:linewidthAvModel}), which represents a scattering of polaritons
into excitonic states. This scattering results in a thermalized population of excitons at high
$\bk$, emitting dominantly from the LP and the bottleneck region, which is the reason for the
observed strong LP emission. Ghost branches are best visible for small $\bkP$ due to the
smaller contribution of the exciton broadening\cite{HuynhPRL03}.

\begin{figure*}
\iffig\includegraphics[width=0.9\textwidth]{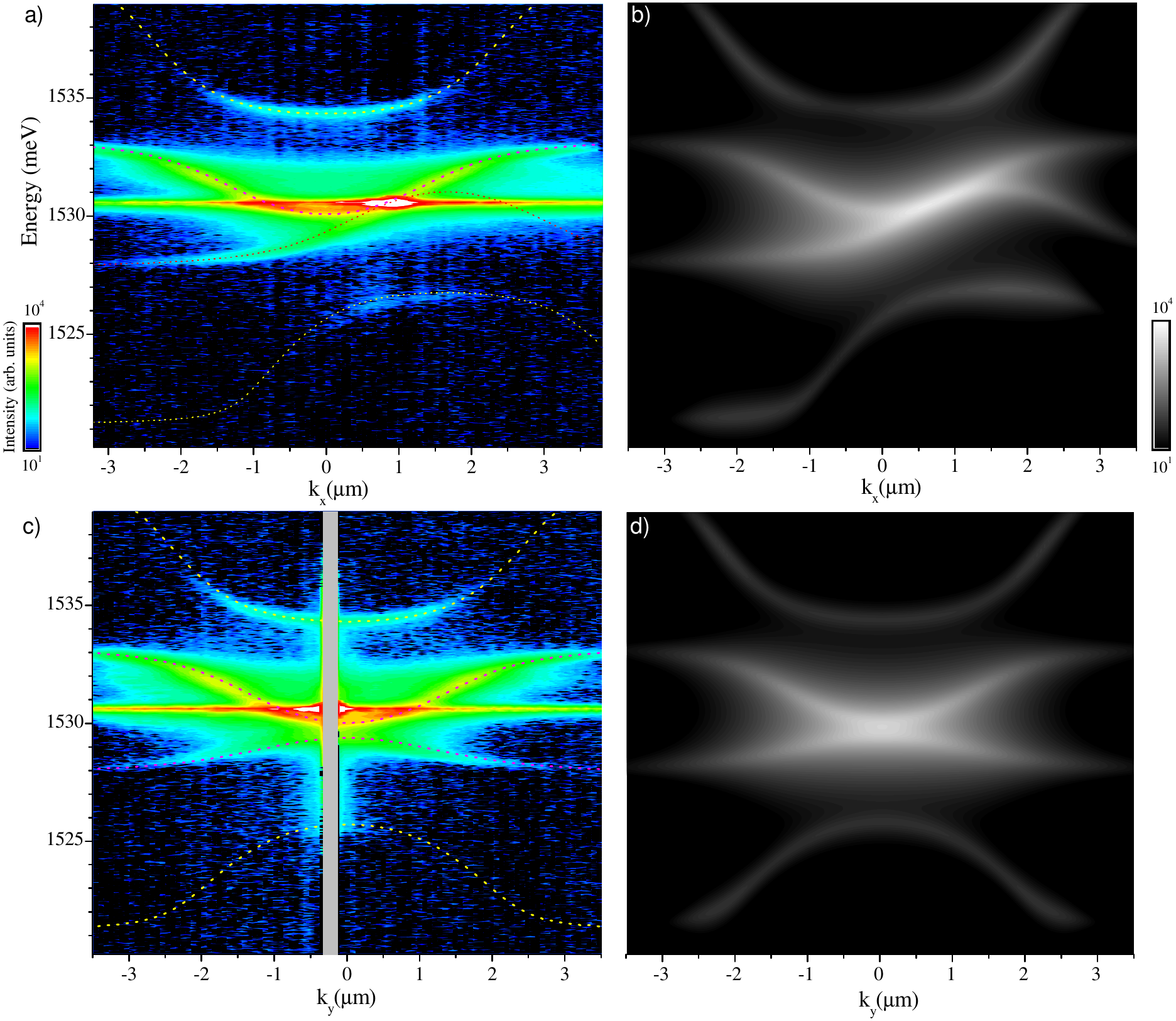} \fi \caption{As
\Fig{fig:D120218012P}, but for $\EP=1530.55$\,meV, $\bkP=(0.85,0)/\mu$m, $\Dc=-2.1$\,meV. a,b)
$\bk=(k_x,0.3/\mu\mbox{m})$. c,d) $\bk=(0/\mu\mbox{m},k_y)$.} \label{fig:D120217008P}
\end{figure*}

Moving the pump away from the dispersion minimum to $\bkP=(0.85,0)/\mu$m, the emission reveals
the expected asymmetry as shown in \Fig{fig:D120217008P}. Two different cross-sections
$\bk=(k_x,0.3/\mu$m) in (a,b) and $\bk=(0/\mu\mbox{m},k_y)$ in (c,d) of the full
three-dimensional data set are given.
%\FloatBarrier
%\FloatBarrier
%
\begin{figure*}
\iffig\includegraphics[width=0.9\textwidth]{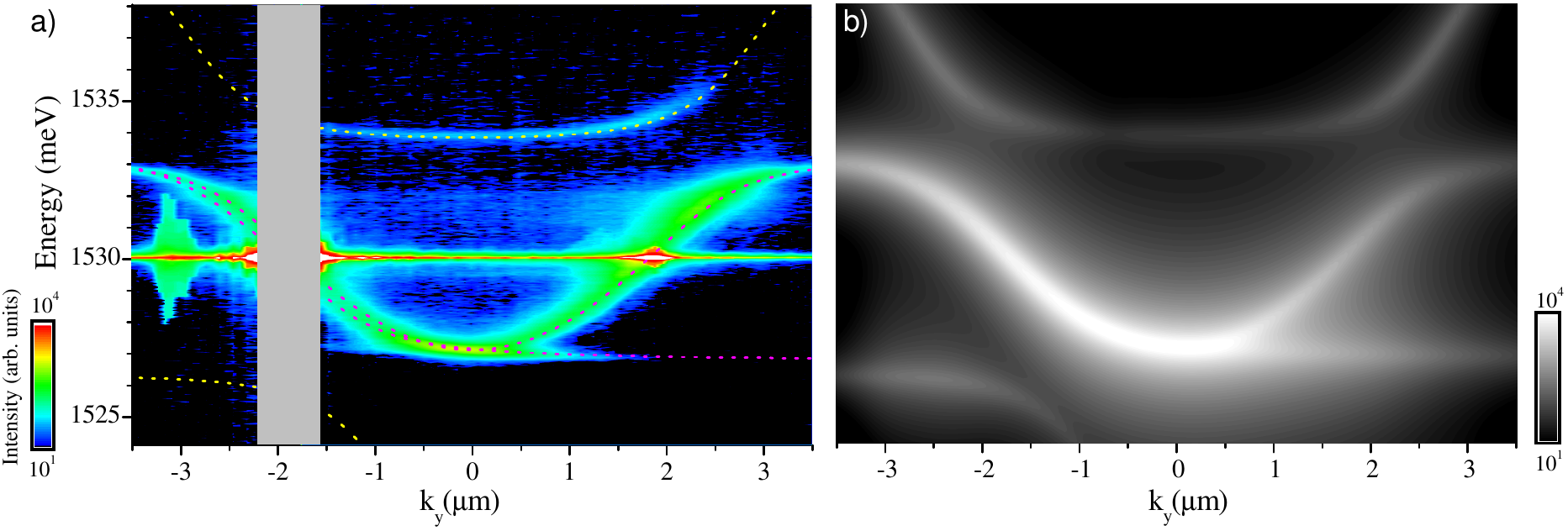} \fi \caption{As
\Fig{fig:D120218012P}, but for $\EP=1530.1$\,meV, $\bkP=(0,-1.9)/\mu$m, $\Dc=-5.5$\,meV, and
cross-section $\bk=(0,k_y)$.} \label{fig:D120216007}
\end{figure*}
Moving further along the dispersion to $\bkP=(0,-1.9)/\mu$m close to the inflexion point, as
shown in \Fig{fig:D120216007}, LP and LP* intersect close to the dispersion minimum at
$\bkP=(0,-0.5)/\mu$m and $\bkP=(0,0.2)/\mu$m, at which energy and momentum conserving
scattering is resonant for signal and idler. This pump wavevector is close to the so-called
magic angle \cite{SavvidisPRL00} for which LP and LP* intersect at $\bk=0$ resulting in a
small threshold for parametric oscillation. For this excitation the ghost branches are visible
mainly at the intersection points. This could partly be due to the onset of stimulated
scattering\cite{LangbeinPRB04a}. The corresponding simulations shown in \Fig{fig:D120216007}b
give good agreement with measurement for the LP branch. However, the calculated MP branch has
a weaker emission for small $\bk$, and a higher emission for large $\bk$. This is again
related to the exciton scattering into the exciton reservoir and subsequent emission of
thermalized excitons, as the middle polariton the highest exciton content at small $\bk$.

\begin{figure*}
\iffig\includegraphics[width=0.9\textwidth]{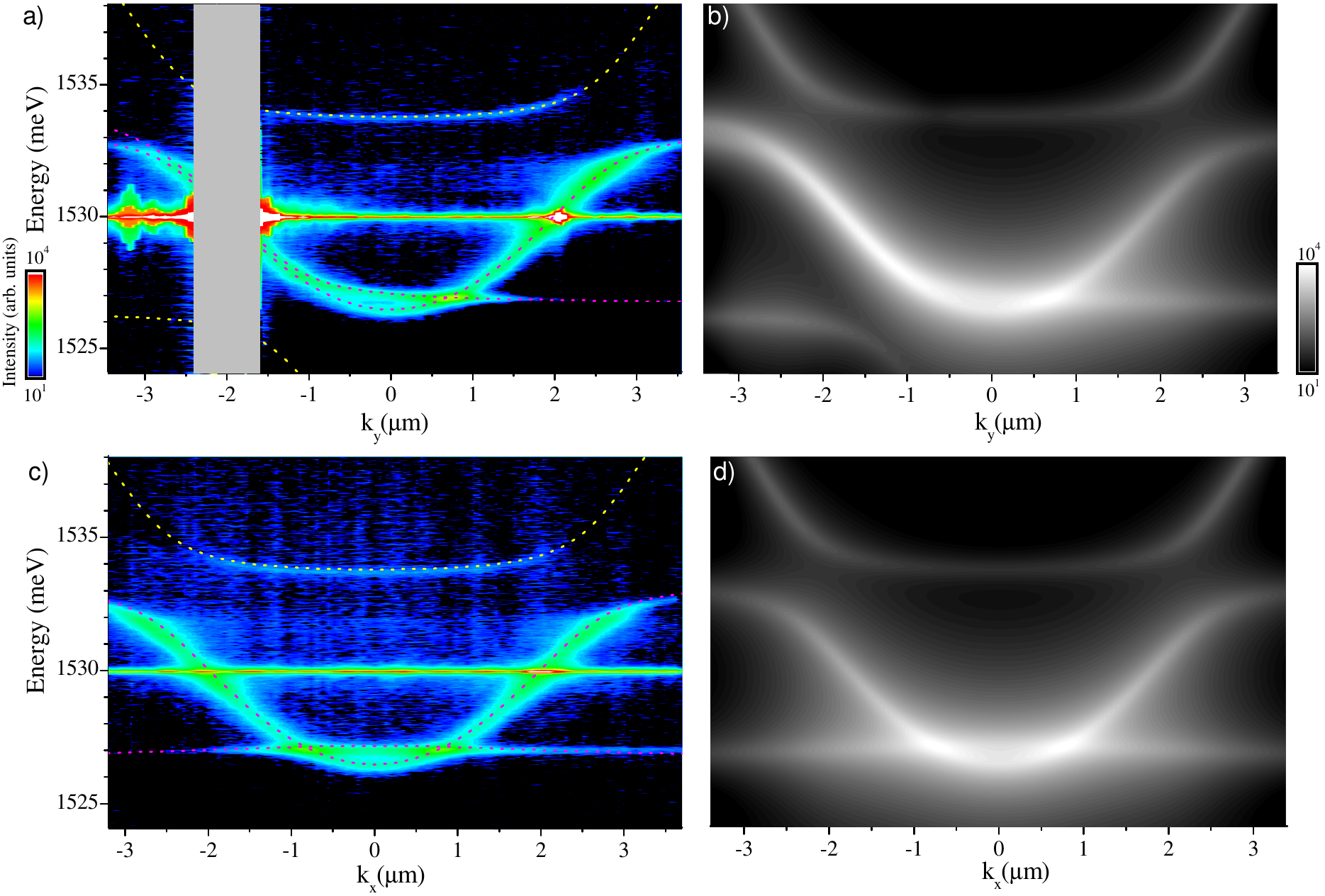} \fi\caption{As
\Fig{fig:D120218012P} but for $\EP=1530.0$\,meV, $\bkP=(0,-1.95)/\mu$m, $\Dc=-6$\,meV, and
$\bk=(-0.25/\mu\mbox{m},k_y)$ for a,b and $\bk=(k_x,0)$ in (c,d),$\delta k =0.14/\mu$m.}
\label{fig:D120216009}
\end{figure*}

In \Fig{fig:D120216009}, we show measured polariton luminescence for $\bkP=(0,-1.95)/\mu$m
well above the inflexion point, resulting in an 8-shaped resonant region\cite{LangbeinPRB04b}
in $\bk$ space. In the cross-section $\bk=(-0.25/\mu\mbox{m},k_y)$, the LP real and ghost
branches intersect at $E=1526.95$\,meV,$\bk=(-0.25,0.7)/\mu$m. In the cross-section
$\bk=(k_x,0)$ shown in \Fig{fig:D120216009}c, the LP real and ghost branches intersect at
$E=1527.1$meV, ${\bf k}=(\pm 0.82,0)/\mu$m. Again a good agreement with the simulations is
found.

%\section{Summary}
%\label{Summary}
In summary we have shown polariton parametric pair scattering from a resonantly excited pump
state into real and ghost branches of signal and idler polaritons for different excitation
angles and wavevectors. The measurements are in agreement with simulations, apart from the
additional emission due to thermalized excitons and the missing treatment of non-radiative
decay. These results can be further explored towards entangled photon source by measurements
of their time-correlation.
\acknowledgments This work was supported by the EPSRC under grant n. EP/F027958/1.

\bibstyle{prbsty}
%\bibliography{JZthesis,langsrv}

\end{document}